# Microwave Rectification by a Carbon Nanotube Schottky Diode

Enrique Cobas and Michael S. Fuhrer
*Department of Physics and Center for Nanophysics and Advanced Materials, University of Maryland, College Park, MD 20742-4111, USA*

**Carbon nanotube Schottky diodes have been fabricated in an all-photolithographic process using dissimilar contact metals on high-frequency compatible substrates (quartz and sapphire). Diodes show near-ideal behavior, and rectify currents of up to 100 nA and at frequencies up to 18 GHz. The voltage and frequency dependence is used to estimate the junction capacitance of ~$10^{-18}$ F and the intrinsic device cut-off frequency of ~400 GHz.**

Due to their high mobility at room temperature[1, 2], semiconducting carbon nanotubes (CNTs) have been proposed for THz electronic devices.[3-7] However, the high characteristic impedance (on order the resistance quantum, $R_Q = h/2e^2 = 12.9$ k$\Omega$) of an individual CNT makes measurements of the high-frequency electrical properties of CNT devices especially difficult because of the impedance mismatch to conventional 50$\Omega$ transmission lines. Several solutions to this problem have been explored, including mixing experiments in three-terminal devices to produce low-frequency output[8, 9], impedance matching using a resonant circuit[10], using multiple CNTs in parallel[11-13] or calculating high-frequency parameters from low-frequency measurements.[14]

We demonstrate a simple solution to the problem of measuring the high-frequency electrical properties of CNT devices: The Schottky diode formed between a CNT and a metal electrode [14-17] is used as a rectifier to produce a DC current in response to a microwave-frequency voltage. We show that this carbon nanotube Schottky diode (CNT-SD) can rectify large currents (> 100 nA) and operates at frequencies up to 18 GHz. We also find that the intrinsic frequency limit is ~400 GHz, and could be much higher if the series resistance of the CNT could be reduced. The Schottky diode can therefore be used as a probe of the local RF power to measure the frequency-dependent conductivity of CNTs and other nanoscale semiconducting devices.

RF-compatible CNT-SD devices were prepared on sapphire and quartz substrates through in the following manner. An array of alumina-supported $Mo_2(acac)_2$ and $Fe(NO_3)_3$ catalyst islands was pre-patterned on sapphire and quartz substrates using a polymethylmethacrylate (PMMA) – photoresist double layer photolithography[18]. Carbon nanotubes (work function 4.7-5.1eV) [19-21] were grown via chemical vapor deposition (CVD) at 900C in a tube furnace using methane, hydrogen and ethylene[22]. Contacts and leads for a 360-device array were created in close proximity to the catalyst islands by two steps of standard photolithography. Low-ohmic contacts were deposited via electron-beam evaporation of 200 Å of platinum (work function 5.6eV), followed by 2,000Å of gold and capped with 200 Å of platinum for surface hardness. Schottky contacts were deposited to either side of the central ohmic contact via thermal evaporation of 50Å of chromium (work function 4.5eV) and 2,000 Å of gold. Figure 1 shows a completed device. The majority of the devices included no nanotubes or included metallic nanotubes which obscure the transport properties of the semiconducting nanotubes (e.g. Device 1 in Figure 1, which exhibits metallic behavior with R~50k$\Omega$). However some of the devices included single semiconducting nanotubes (e.g. Device 2 in Figure 1).

All electrical measurements were carried out in a Cascade Microtech Summit 1200 automated probe station using Keithley 2400 Sourcemeters under ambient conditions; see Supplementary Information for more details about the measurement setup. RF measurements were performed using a Hewlett Packard 83620B microwave signal generator, an X-Band horn antenna, an Ithaco 1211 current amplifier and a Stanford Research Systems SR 380 lock-in amplifier. The probe station's dc electrical probes doubled as receiver antennae, directing the microwave signal broadcast from the nearby horn antenna into the nanotube device. The microwave signal was amplitude modulated at 1 kHz to enable measurement of the rectified current signal with the lock-in amplifier as a function of microwave frequency, microwave power and dc bias.

Figure 2 shows the dc current-voltage (*I-V*) curve for an individual CNT-SD. At low frequency,

we model the CNT-SD as a diode in series with a resistor:

$$I = I_s \left[ e^{(qV-IR_s)/nkT} - 1 \right], \quad (1)$$

where $I_s$ is the reverse saturation current, $R_s$ is the series resistance, $n$ is the ideality factor, $k$ is Boltzmann's constant, $q$ is the electronic charge, and $T$ is the temperature. A fit to this equation (0mV to 400mV) gives $n = 1.00$ and a series resistance $R_s = 420$ k$\Omega$.

The diode resistance and series resistance are approximately equal at $V_{dc} \approx 200$mV, where a knee in the $I$-$V$ curve is evident.

In general we found that the response of the device embedded in the measurement setup was strongly dependent on frequency, with the response as a function of frequency showing numerous peaks and valleys. We attribute this to strong dependence of the antenna efficiency on frequency resulting from resonances in our non-ideal setup. Thus using the applied microwave power is a poor indicator of the power reaching the device. Although the complex frequency dependence of our measurement setup makes a simple, direct frequency dependence measurement impossible, a comparison of the bias voltage dependence of the data at various frequencies within the square-law power regime does lead us to useful conclusions. Figure 3 shows a comparison of rectified current $\Delta I$ at frequencies $f = 7$ GHz (red) and $f = 18$ GHz (blue) at various source powers $P$ within the square-law regime. A systematic difference in the shape of the $\Delta I$-$V$ curve with frequency is evident. This may be understood qualitatively as follows. Consider the two curves corresponding to $f = 7$ GHz, $P = -14$ dBm and $f = 18$ GHz $P = -8$ dBm. At high $V$, when the diode resistance is low, these two curves produce similar response, indicating that the cutoff frequency is much larger than 18 GHz. As $V$ is decreased, the diode resistance increases, and the cutoff frequency decreases below 18 GHz so the response becomes frequency-dependent, smaller at higher $f$. We explore this quantitatively below.

We express the power reaching the Schottky junction as

$$P_j = P E_{ant} E_{dev} \quad (2)$$

where $P$ is the emitted power, $E_{ant}$ is the fraction of the emitted power absorbed by the device and $E_{dev}$ is the fraction of power in the device dissipated at the Schottky junction. Then the additional current due to the power absorbed in the junction can be expressed as

$$\Delta I = \beta P_j \quad (3)$$

where the responsivity $\beta$ is determined solely by the variable junction resistance $R_j(V)$.

Then, following Sorensen et al,[23] we model the device as $R_j$ in parallel with a constant junction capacitance $C_j$ and a constant series resistance $R_s$, as in Figure 3 (inset). This leads to expressions for the predicted cut-off frequency $f_c$ in the square-law regime and the frequency-dependent device efficiency $E_{dev}(f)$:

$$f_c = \frac{[1 + R_s/R_j]^{\frac{1}{2}}}{2\pi C_j (R_s R_j)^{\frac{1}{2}}} \quad (4)$$

$$E_{dev}(f) = \frac{P_j}{P_{dev}} = \frac{1}{[1 + \frac{R_s}{R_j}][1 + (f/f_c)^2]} \quad (5)$$

then from Eqn. 2 the ratio of the currents at two different frequencies is $\Delta I(f_1)/\Delta I(f_2) = [E_{dev}(f_1)/E_{dev}(f_2)] \times [E_{dev}(f_1)/E_{dev}(f_2)] = $ const. x $[E_{dev}(f_1)/E_{dev}(f_2)]$. The signal strength was verified to be proportional to the microwave power, indicating the device was operating in the square-law regime. The ratio of the device efficiencies at frequencies $f_1$ and $f_2$ approaches the values $E_{dev}(f_1)/E_{dev}(f_2) = 1.00$ for $(f_1, f_2 \ll f_c)$ and $E_{dev}(f_1)/E_{dev}(f_2) = (f_2/f_1)^2$ (which is a factor of 6.61 for the two frequencies used here) for $(f_1, f_2 \gg f_c)$. Since $R_j$ and hence $f_c$ are strong functions of bias voltage, in general there will be a transition from $f_1, f_2 \gg f_c$ to $f_1, f_2 \ll f_c$, and the ratio $\Delta I(f_1)/\Delta I(f_2)$ will be bias-voltage dependent.

Figure 4 shows a comparison of the efficiency ratio for various trial junction capacitances $C_J$ and the ratio of the experimentally measured 7 GHz and 18 GHz signals, $\Delta I(7$ GHz$)/\Delta I(18$ GHz$)$, which, as anticipated above, is voltage dependent. The experimental ratio varies more gradually with voltage than the theoretical curves. This is likely due to the fact that $R_s$ and $C_j$ are distributed elements along the length of the CNT; as $R_j$ becomes small (large *V*) and the resistance is dominated by the series resistance of the CNT, the capacitive coupling of portions of the CNT more distant from the junction becomes relatively more important, and the effective $C_j$ increases. However, comparing the large *V* and small *V* limits we can conclude that the junction capacitance of the device is in the $10^{-18}$ F range, in accordance with previous estimates[24-26]. This value allows us to estimate the intrinsic nanotube diode cut-off frequency

$$f_c = \frac{1}{2\pi RC_j} = \frac{1}{2\pi (420 \text{ k}\Omega)(10^{-18} \text{ F})} \approx 400 GHz$$

We have demonstrated fabrication of carbon nanotube Schottky diodes via straightforward CNT synthesis and lithography techniques. The devices produced exhibit excellent Schottky diode characteristics with ideality factors *n* close to unity, in series with a high channel resistance and exhibit good rectification performance beyond 7GHz. We have also demonstrated a simple technique for observation of microwave rectification in nanoscale semiconductor devices. The intrinsic cut-off frequency set by the junction capacitance and series resistance of CNT-SDs is on order 400 GHz and could be pushed into the THz if the series resistance could be lowered, e.g. through shortening or doping the CNT. Measurements on an additional CNT-SD (see Supplementary Information) indicate that the series resistance can be reduced by shortening the CNT.

We acknowledge support from the Army Research Laboratory MICRA program, the shared equipment facilities of the UMD-MRSEC and thank Dr. Steven Anlage for useful discussions and use of his microwave equipment.

1. T. Durkop, S. Getty, E. Cobas and M. S. Fuhrer, Nano Letters 4, 35 (2004).
2. V. Perebeinos, J. Tersoff and P. Avouris, Nano Letters 6, 205 (2005).
3. P. J. Burke, Solid State Electronics 48, 1981 (2004).
4. S. H. Jing Guo, A. Javey, G. Bosman and M. Lundstrom IEEE Transactions on Nanotechnology 4, 715 (2005).
5. K. Alam and R. K. Lake, Journal of Applied Physics 100, 024317 (2006).
6. L. C. Castro, D. L. Pulfrey, and D. L. John, Solid State Phenomena Part B. 121-123, 693 (2007).
7. D. Dragoman and M. Dragoman, Physica E 25, 492 (2005).
8. A. A. Pesetski, J. E. Baumgardner, E. Folk, J. Przybysz, J. Adam, H. Zhang, Applied Physics Letters 88, 113103 (2006).
9. S. Rosenblatt, H. Lin, V. Sazonova, S. Tiwari and P. McEuen., Applied Physics Letters 87, 153111 (2005).
10. S. Li, Z. Yu, S.-F. Yen, W. C. Tang and P. J. Burke, Nano Letters 4, 753 (2004).
11. X. Huo, M. Zhang, P. C. H. Chan, Q. Liang and Z. K. Tang., International Electron Devices Meeting 2004, 691 (2005).
12. J. M. Bethoux, Happy, H., G. Dambrine, V. Derycke, M. Goffman and J. P. Bourgoin, IEEE Electron Device Letters 27, 681 (2006).
13. T.-I. Jeon, K.-J. Kim, C. Kang, I. H. Maeng, J.-H. Son, K.H. Son, K. H. An, J. Y. Lee, and Y. H. Lee, Journal of Applied Physics 95, 5736 (2004).
14. H. Manohara, E. Wong, E. Schlect, B. D. Hunt and P. H. Siegel., Nano Letters 5, 1469, (2005).
15. S. Heinze, J. Tersoff, R. Martel, V. Derycke, J. Appenzeller and Ph. Avouris., Physical Review Letters 89 (2002).
16. F. Leonard and J. Tersoff, Physical Review Letters 84, 4693 (2000).
17. W. Zhu and E. Kaxiras, Applied Physics Letters 89, 243107 (2006).
18. J. Kong, H. Soh, A. Cassell, C. Quate and H. Dai, Nature 395, 878 (1998).
19. S. Suzuki, Y. Watanabe, Y. Homma, S. Fukuba, S. Heun and A. Locatelli, Applied Physics Letters 85, 127 (2004).
20. S. Suzuki, C. Bower, Y. Watanabe and O. Zhou, Applied Physics Letters 76, 4007 (2000).
21. K. Okazaki, Y. Nakato and K. Murakoshi, Physical Review B 68, 035434 (2003).
22. W. Kim, H. C. Choi, M. Shim, Y. Li, D. Wang and H. Dai, Nano Letters 2, 703 (2002).
23. H. O. Sorensen and A. M. Cowley, IEE Transactions on Microwave Theory and Techniques MTT-14, 588 (1966).
24. P. Jarillo-Herrero, S. Sapmaz, C. Dekker, L. P. Kowenhoven and H. S. H. J. van der Zant, Nature 429, 389 (2004).
25. H. W. C. Postma, T. Teepen, Z. Yao, M. Grifoni and C. Dekker., Science 293, 76 (2001).
26. R. Tarkiainen, M. Ahlskog, J. Penttila, L. Roschier, P. Hakonen, M. Paalanen and E. Sonin, Physical Review B 64, 195412 (2001).

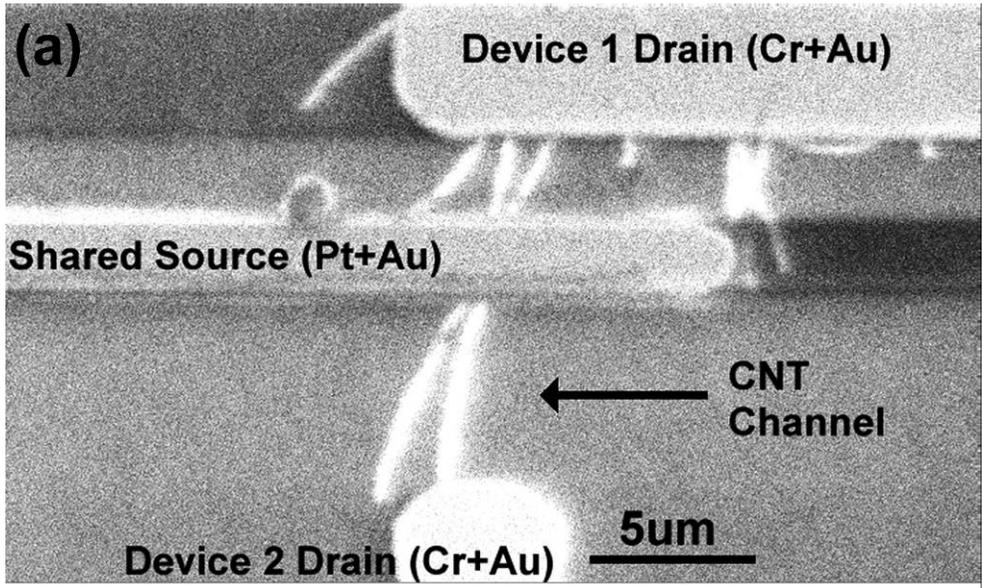

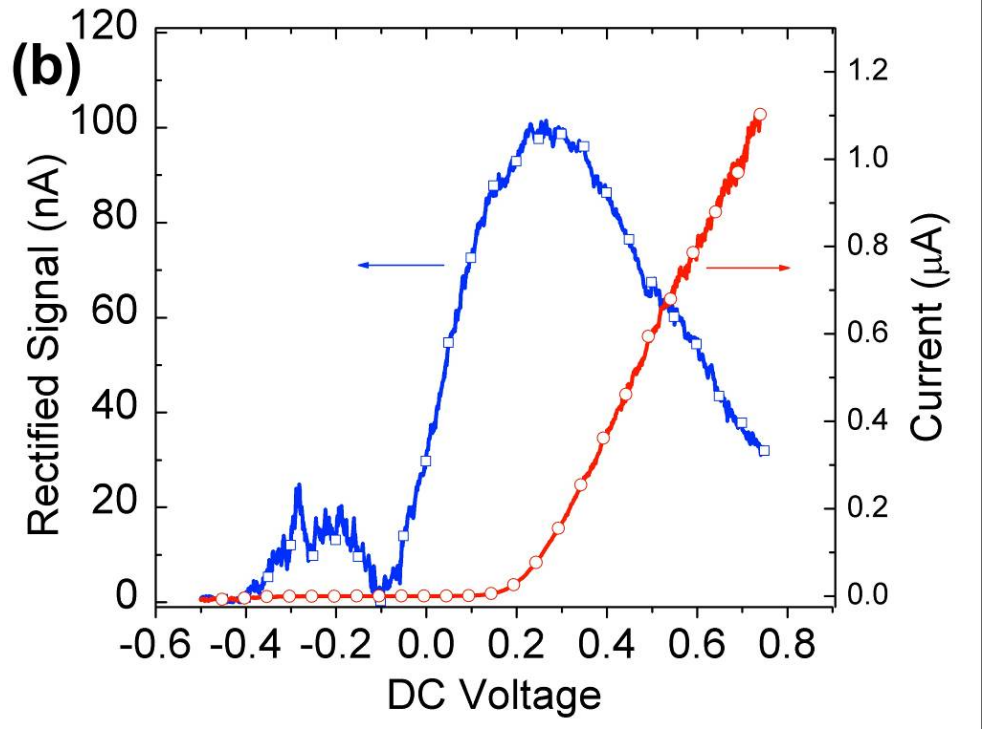

**Figure 1.** (a) Optical image of the leads comprising devices 1 and 2, and (b) the current-voltage characteristics (red circles) and rectified current signal (blue squares) from Device 2 shown, due to a 7GHz microwave signal at 9dBm.

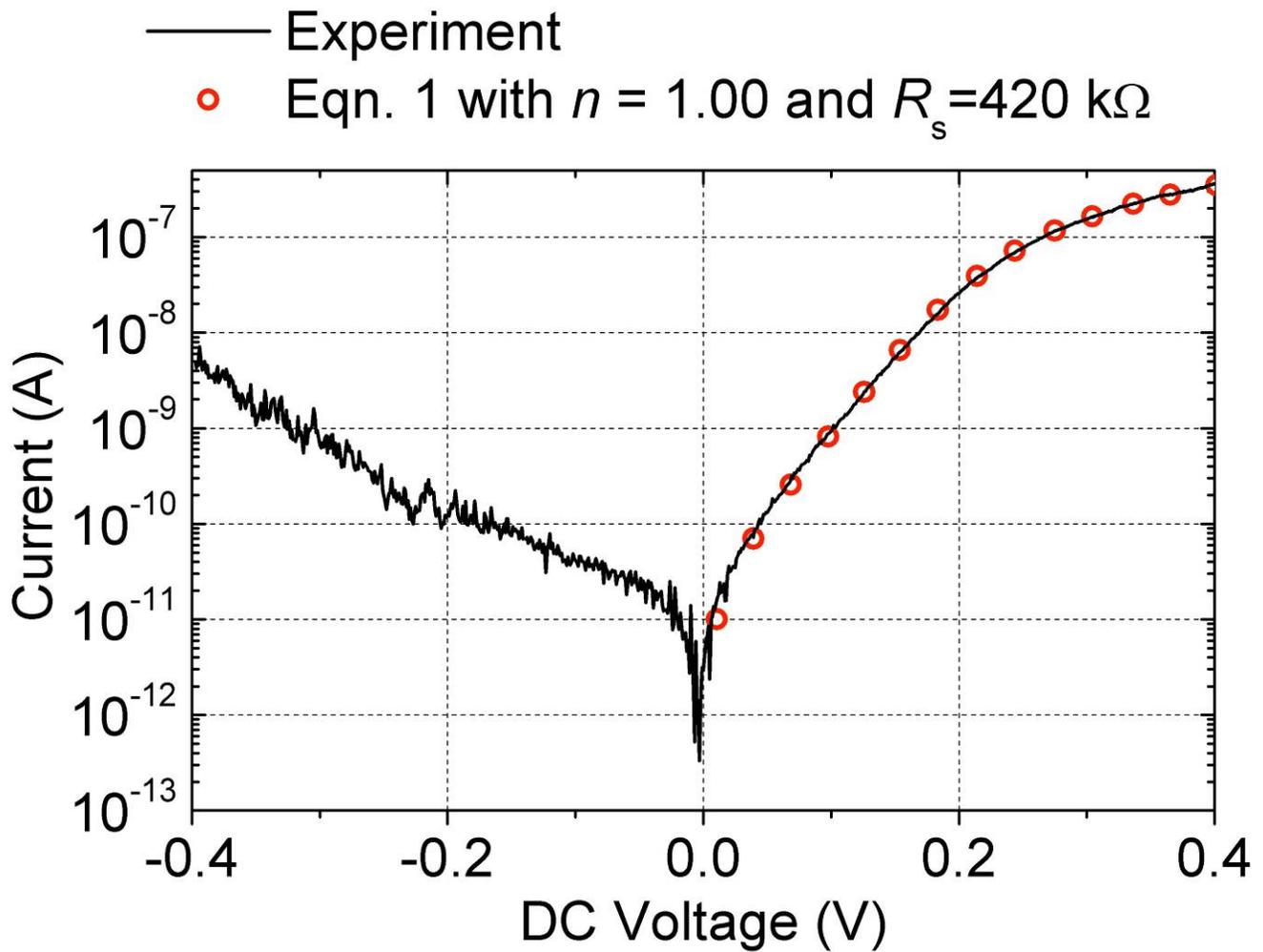

**Figure 2**. DC current-voltage characteristics of a CNT Schottky diode (black line) and corresponding fit to Eqn. 1 (red circles) with *n* = 1.00 and $R_s$ = 420kΩ. The data shown is from Device 2 pictured in Figure 1.

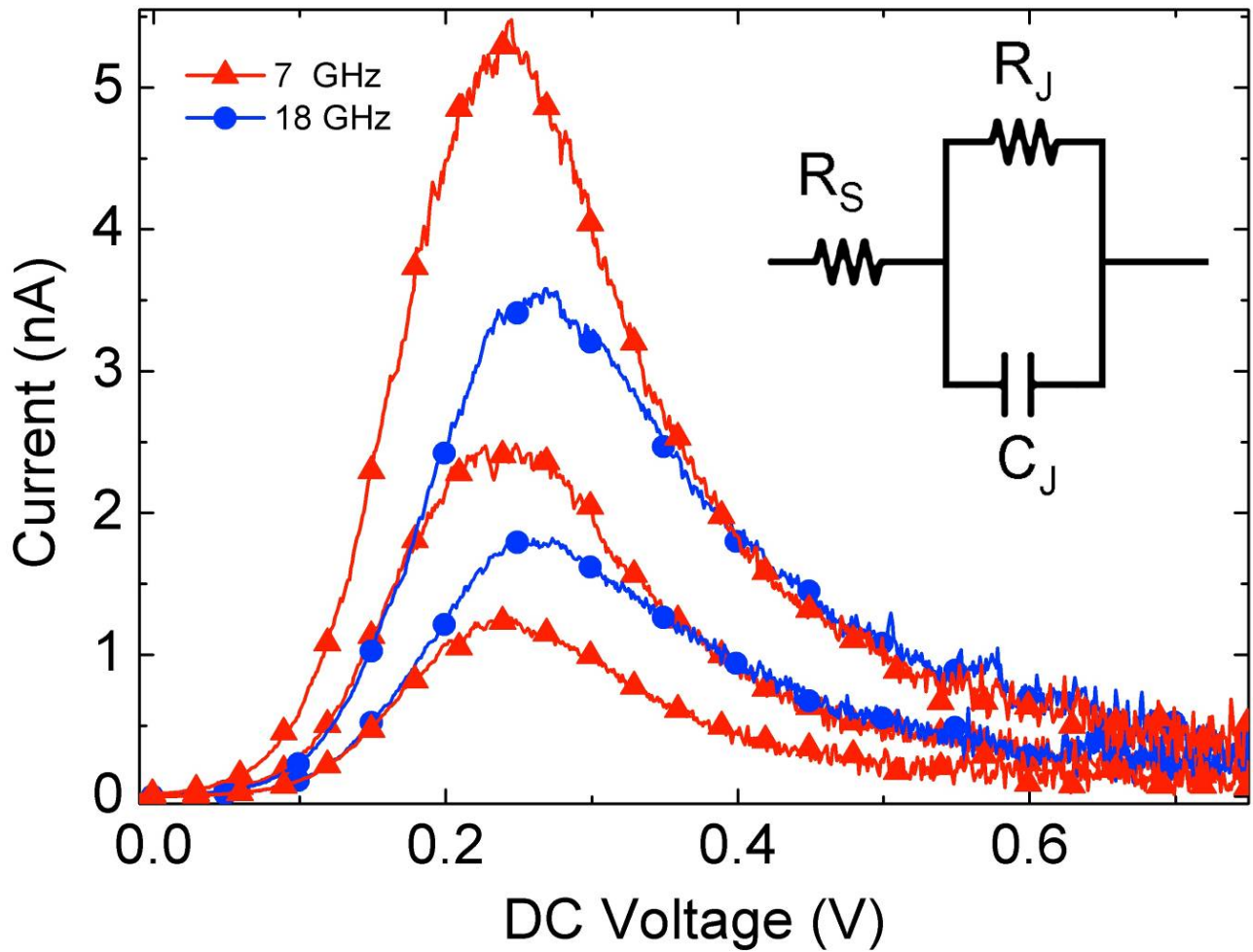

**Figure 3.** Rectified output signal from a CNT-SD as a function of dc bias under various powers of 7GHz (red squares) and 18GHz (blue triangles) microwave excitation. The corresponding microwave intensities are (top to bottom) -14dBm, -17dBm and -20dBm for the 7GHz data and -8dBm and -11dBm for the 18GHz data. Inset shows the model equivalent circuit, ignoring the stray capacitance which is bias-independent.

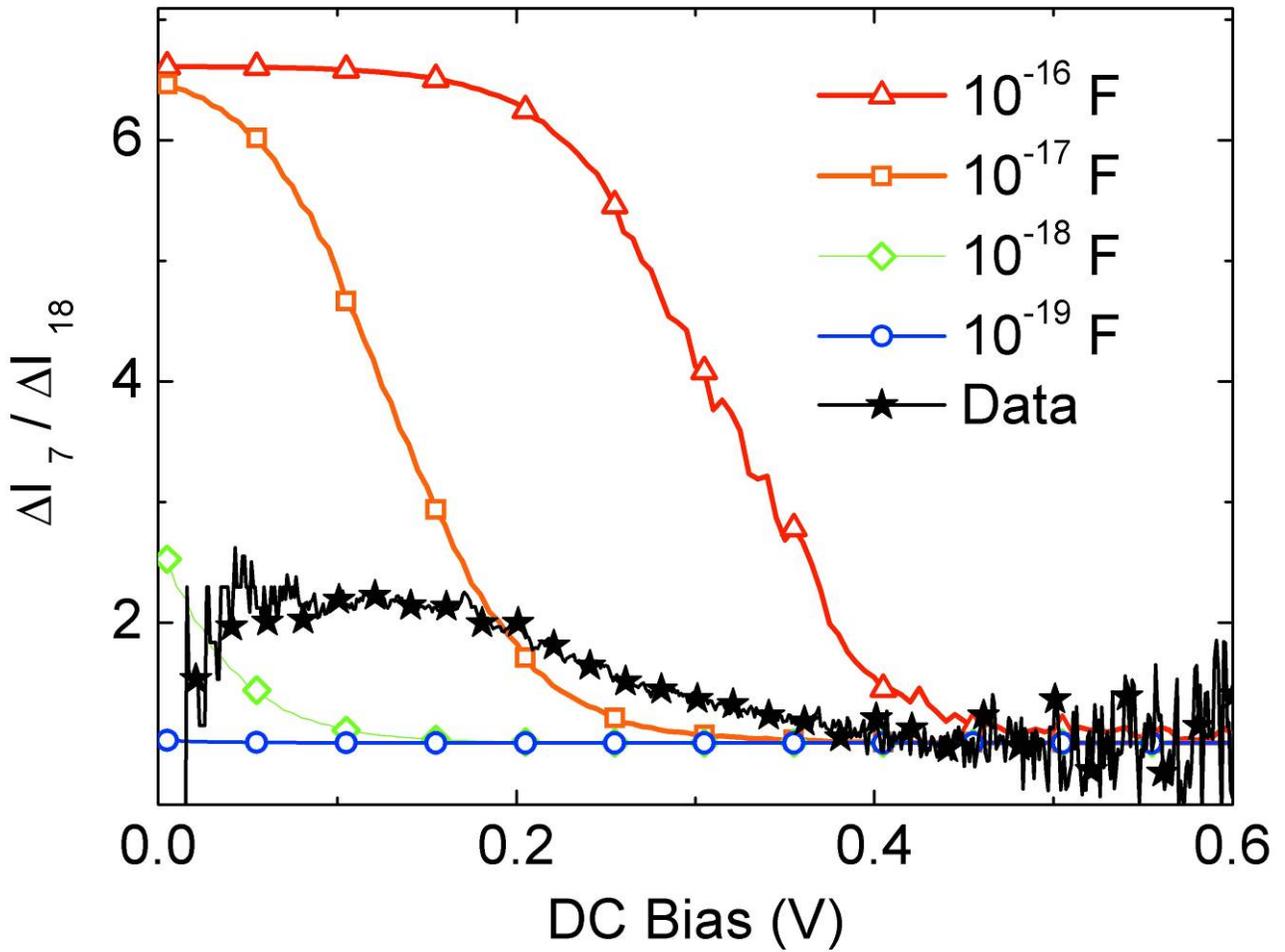

**Figure 4**. Rectified current ratio at 7GHz vs.18GHz as a function of dc bias. Solid lines are calculated from the measured junction resistance data $R_j(V)$ and the model equivalent circuit in figure 3 for junction capacitances $C_j = 10^{-19}$ F (blue circles), $10^{-18}$ F (green rhombi), $10^{-17}$ F (orange squares), $10^{-16}$ F (red triangles). The experimentally observed ratio (scaled to 1.0 at high bias) is shown as black stars.

Supplementary Information for E. Cobas and M. S. Fuhrer, "Microwave Rectification by a Carbon Nanotube Schottky Diode"

I. Device Geometry and Measurement Setup

Figures S1 and S2 give additional details about the macroscopic setup used to measure the carbon nanotube Schottky diodes (CNT-SDs). Figure S1 shows the arrangement of lithographically-patterned leads and the probe electrodes used to make contact to the lithographic leads. Figure S2 shows the experimental setup including device under test, probes, horn antenna, and optical microscope. The setup is complex on the scale of the wavelength of radiation (1.7 - 4.3 cm) being detected, and it is therefore reasonable that the frequency response is complex and shows numerous resonances.

II. Additional diode

Figure S3 shows the I-V characteristics of an additional CNT-SD (Device 3). The CNT in Device 3 which had an active length $L = 3$ μm, and a fit to the diode equation (Eqn. 1 in main text) gives a series resistance $R_s = 228$ kΩ. Comparing this to Device 2 discussed in text ($L = 6$μm, $R_s = 420$ kΩ), it is evident that decreasing $L$ can be used to decrease the series resistance of the device, and increase the cutoff frequency.

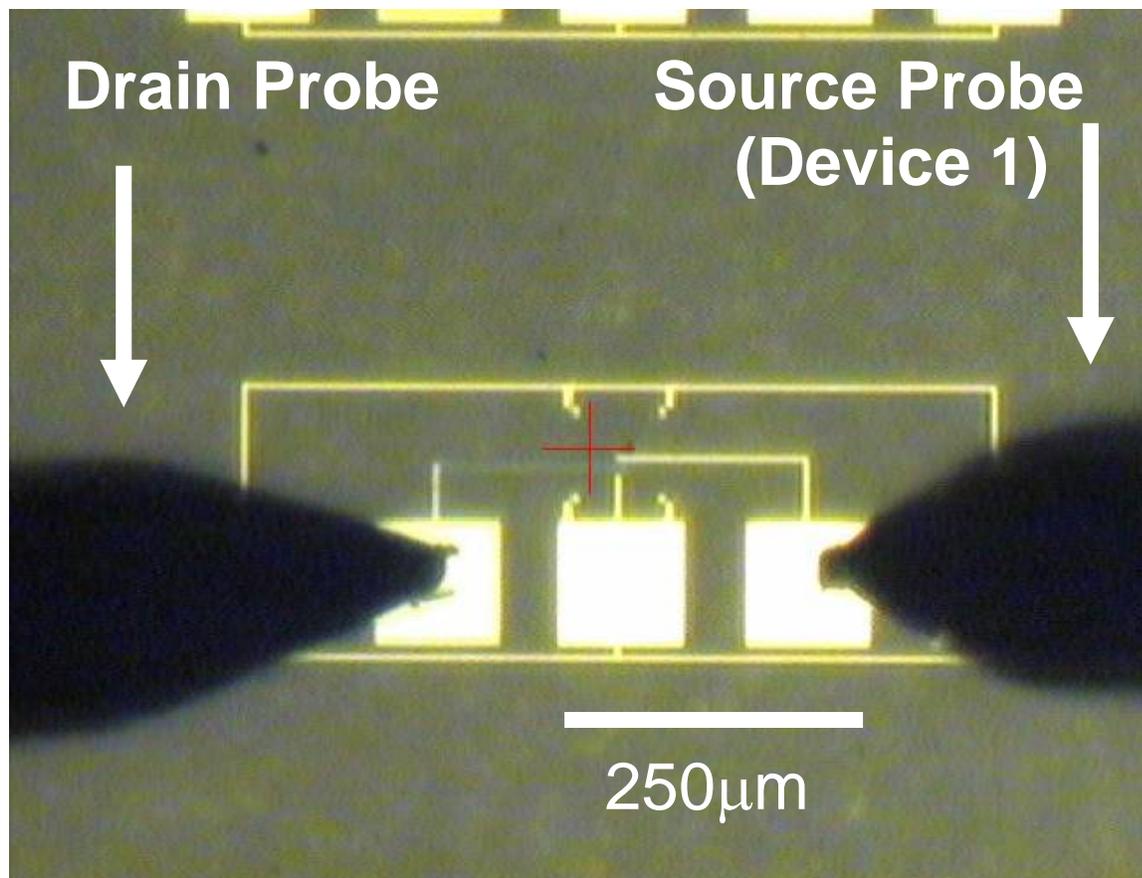

Figure S1. Optical image of an electrode set (Devices 3 and 4) and source/drain probes during microwave measurement of Device 3. The microwave antenna was positioned orthogonal to the probes axis (top, not shown) with the polarization oriented in-plane with the probes. Another electrode set can be seen at the top edge of the image.

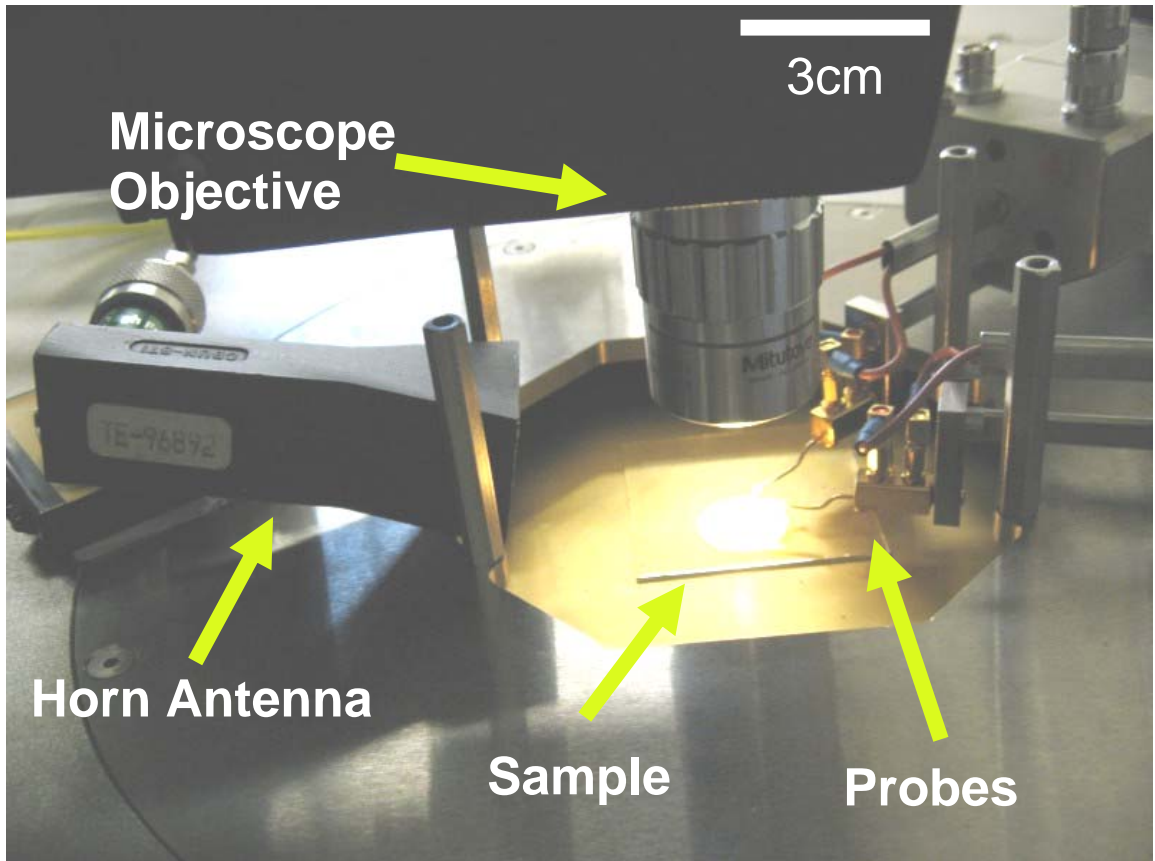

Figure S2. Measurement setup including the X-band horn antenna, sample, electrical probes and microscope lens in addition to the probe station stage. (Note that the probes are not positioned for microwave measurements in this photo, and the sample shown here is an array of commercially available GaAs Shottky diodes from Arotech Inc. That data was not used.)

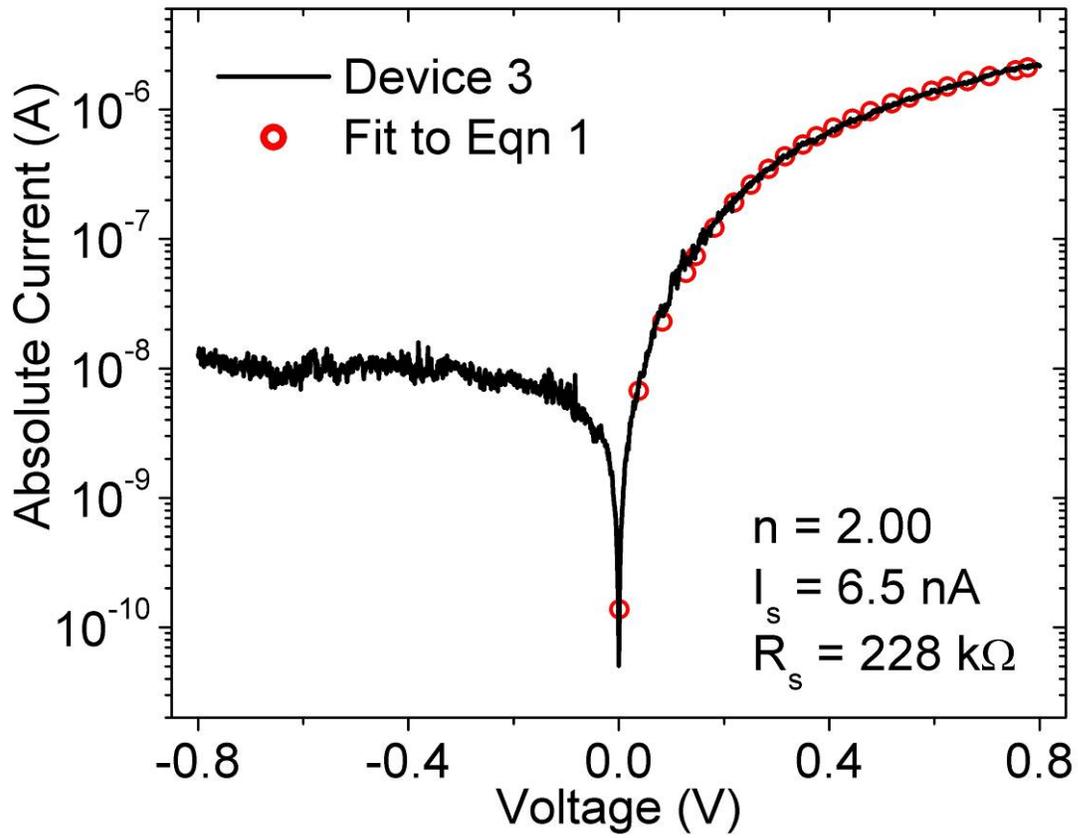

Figure S3. Current-voltage characteristics for Device 3 (black line) and a corresponding fit to the diode equation, Eqn 1. (red circles) with $n$ = 2.00 and $R_s$ = 228kΩ. Channel length $L$ for Device 3 is ~3 microns.